\documentclass[twocolumn,prl,english,superscriptaddress,floatfix,preprintnumbers,showpacs]{revtex4-2}
\usepackage[T1]{fontenc}
\usepackage{babel}
\usepackage{verbatim}
\usepackage{amsmath}
\usepackage{amssymb}
\usepackage{graphicx}
\usepackage[unicode=true,pdfusetitle,
bookmarks=true,bookmarksnumbered=false,bookmarksopen=false,
 breaklinks=false,pdfborder={0 0 1},backref=false,colorlinks=false]
 {hyperref}

\begin{document}
\title{Bosonization of the interacting Su-Schrieffer-Heeger model}
\author{Tony Jin}
\email{tonyjin@uchicago.edu.ch}
\affiliation{DQMP, University of Geneva, Quai Ernest-Ansermet 24, CH-1211 Geneva, Switzerland}
\affiliation{Pritzker School of Molecular Engineering, University of Chicago, Chicago, Illinois 60637, USA}
\author{Paola Ruggiero}
\affiliation{King's College London, Strand, WC2R 2LS London, United Kingdom}
\author{Thierry Giamarchi}
\affiliation{DQMP, University of Geneva, Quai Ernest-Ansermet 24, CH-1211 Geneva, Switzerland}

\begin{abstract}
We derive the bosonization of the interacting fermionic Su-Schrieffer-Heeger (SSH) with open boundaries. We use the classical Euler-Lagrange equations of motions of the bosonized theory to compute the density profile of the Majorana edge mode and observe excellent agreement with numerical
results, notably the localization of the mode  near the boundaries. Remarkably, we find that repulsive or attractive interactions do not systematically localize or delocalize the edge mode but their effects depend on the value of the staggering parameter. We provide quantitative predictions of these effects on the localization length of the edge mode. 
\end{abstract}
\maketitle
Topological concepts have become a central part of contemporary condensed
matter physics \cite{Haldane_NobelLecture}. The understanding of the geometrical and topological
objects underpinning band theory \cite{AltlandZirnbauer_tenfold,ReviewTenfoldwayShinseiRyu,NakaharaBook,Cayssol_FuchsIntroductiontotopologybandtheory} has fostered intense activities in diverse areas of condensed matter physics such
as the study of the quantum Hall effect \cite{KlitzingIntegerQHalleffect,40yearsHalleffect}, spin-orbit induced topological
band insulators \cite{LiangKaneTopologicalinsulator,KaneMeleeSpinHallGraphene,BernevigQuantumspinHall}, topological quantum computing \cite{ReviewAnyonstopologicalQcomputation}  and so on and so forth. 

One current limitation of topological band theory is its
restriction to non-interacting systems. Interactions
will in general spoil the band structure, rendering usual classification
schemes inoperative. There, novel phenomena may be expected, the most
famous example being the fractional quantum Hall effect \cite{FractionalQHELaughlin}. 

One of the simplest model capturing the key features of topological
insulators is the Su-Shrieffer-Heeger (SSH) model \cite{SSHoriginalpaper}. The fermionic
SSH model consists of a 1D tight-binding model with alternating bond
value. Depending on whether the first bond is weak (strong) the model
is either in the topological (trivial) phase. For open boundaries,
one possible characterization of the topological phase is the presence 
of two-fold, quasi-degenerate, zero energy, Majorana edge modes that are exponentially
localized at the boundaries. 
Remarkably, these edge modes have been observed and characterized experimentally in one-dimensional optical lattices \cite{Bloch_TopologicalPhase} and artificial spin chains simulated with Rydberg atoms \cite{DeLeseleuc_ExpRydbergSSH} in ultracold atoms setup. 

Although the SSH is now considered a textbook model for topological
insulators, the inclusion of interactions in this model remains to
this day an open question. On the other hand, a powerful technique
that was developped in the previous decades to treat interacting fermionic
systems in 1D is \emph{bosonization} \cite{Haldane_Bosonization,Thebook}. 
One trademark prowess of bosonization is to
map interacting spinless fermions in 1D to a \emph{free} bosonic theory. Remarkably this technique has been successfully applied to study the effects of interactions on Majorana modes in superconducting wires \cite{SimonLoss_BosonizationMajorana,Sarma_BosonizationMajorana,Loss_Majorana0modes} but so far, to the best of our knowledge, the SSH model has escaped from a similar treatment. In particular, the spatial localization of the edge modes has not been described within the bosonization language.

In this paper, we fill this gap by deriving
the bosonized theory of the interacting SSH model with open boundaries. We use the classical Euler-Lagrange (EL) equations of motion of the bosonized theory to compute the density profile of the Majorana edge mode and observe excellent agreement with numerical
results, notably the localization of the mode near the boundaries. Remarkably, we find that repulsive or attractive interactions do not systematically localize or delocalize the edge mode but their effects depend on the value of the staggering parameter. We provide quantitative predictions of these effects on the localization length. Our results pave the way to a generalization to other interacting topological models. 

We begin by discussing the bosonization procedure in the absence of
interactions. Let $(c_{j},c_{j}^{\dagger})_{j\in[1,N]}$ be the usual
fermionic creation and annihilation operators associated to site $j$.
The discrete free SSH Hamiltonian on $N$ sites in 1D is given by
\begin{equation}
H=\sum_{j=1}^{N-1}\left(-t-\delta(-1)^{j}\right)\left(c_{j}^{\dagger}c_{j+1}+{\rm h.c}\right), \label{eq:discreteHamiltonian}
\end{equation}
i.e we have a tight binding chain with alternating values for the
bond. Fixing $t>0$, the topological phase corresponds to the case
where we have an \emph{even }number of sites and $\delta>0$. A signature
of the topological phase is the existence for open conditions of quasi-degenerate zero energy eigenstates \footnote{for a finite system, the two modes are, strictly speaking, degenerate but their energy approaches $0$ exponentially fast as one increases the system size.} in which a single particle is in
a coherent superposition between the two edges of the chain
- see e.g \cite{AsbothBookintroductiontopologicalinsulator,CoursDalibardTopo} for details on the non-interacting case. 

The Fourier transform for open boundaries is given by 
\begin{equation}
c_{j}=i\sqrt{\frac{2}{N+1}}\sum_{n=1}^{N}\tilde{c}_{n}\sin\left(\frac{\pi jn}{N+1}\right).\label{eq:discrete FT}
\end{equation}
For $\delta=0$, this rotation diagonalizes the problem, i.e we have
$H_{\delta=0}=\sum_{n}\varepsilon_{n}\tilde{c}_{n}^{\dagger}\tilde{c}_{n}$
with $\varepsilon_{n} \equiv  -2t\cos\left(\frac{\pi n}{N+1}\right)$.

The continuum limit is obtained by introducing the lattice spacing
$a$ and defining the position $x \equiv  ja$ and the momentum $k \equiv  \frac{2\pi n}{2a(N+1)}$.
The size of the sytem is taken to be $L \equiv  a(N+1)$ so that $k=\frac{\pi n}{L}$.
The continuous fermionic field is $\Psi(x) \equiv  \frac{1}{\sqrt{a}}c_{j}$.
The boundary conditions for $\Psi$ are obtained by extending the
discrete formula (\ref{eq:discrete FT}) to site $0$ and site $N+1$
:\emph{ }$\Psi(0)=0$ and $\Psi(x=a(N+1)=L)=0$. 

Following the usual bosonization procedure \cite{Thebook,DelfSchoeller_Bosonization}, we split $\Psi$
into a left and a right moving fields $\Psi_{{\rm R/L}}$ by expanding
around the Fermi energy $\varepsilon_{k_{F}}$: 
\begin{align}
\Psi(x) & =\Psi_{{\rm R}}(x)+\Psi_{{\rm L}}(x),\\
\Psi_{{\rm R}}(x) & =\sqrt{\frac{L}{2}}e^{ik_{F}x}\int_{-\infty}^{\infty}\frac{dk}{\pi}\tilde{c}_{k+k_{F}}e^{ikx},\\
\Psi_{{\rm L}}(x) & =-\Psi_{{\rm R}}(-x).\label{eq:relationleftright}
\end{align}
For convenience, we also define the ``slow'' fields ${\rm R}(x) \equiv  \Psi_{{\rm R}}(x)e^{-ik_{F}x},\quad{\rm L}(x) \equiv  e^{ik_{F}x}\Psi_{{\rm L}}(x).$
Importantly, because of open boundaries, the left and right movers
are \emph{not }independent as is encapsulated by Eq.(\ref{eq:discrete FT}), see e.g \cite{FabrizioGogolin_Bosonizationopen,Johanesson_Bosonizationopen,Cazalilla_Bosonization1D} for previous discussions of open boudaries bosonization.

In the continuum, ${\rm R}(x)$ taken alone can be thought of as a
field living on a space of size $2L$ with \emph{periodic boundary
conditions}. In the bosonized language, it can then be reexpressed
as 
\begin{align}
{\rm R}(x) & =F_{{\rm R}}\frac{1}{\sqrt{2\pi\alpha}}e^{i\phi^{R}(x)},\\
\phi^{{\rm R}}(x) &  \equiv  \frac{\pi x}{L}N_{{\rm R}}+\sum_{n>0}\frac{1}{\sqrt{n}}(a_{n}e^{i\frac{n\pi x}{L}}+a_{n}^{\dagger}e^{-i\frac{n\pi x}{L}})e^{-\alpha\frac{\pi n}{2L}}.
\end{align}
where $F_{{\rm R}}$ is the Klein factor associated to the right-mover
with $F_{{\rm R}}^{\dagger}F_{{\rm R}}=F_{{\rm R}}F_{{\rm R}}^{\dagger}=1$,
$N_{{\rm R}}$ the particle number operator associated to the right
movers, $(a_{n},a_{n}^{\dagger})_{n>0}$ a set of bosonic modes indexed
by $n$ and $\alpha$ a regularization parameter. The expression of
the bosonic field associated to the left-movers can be readily deduced
from (\ref{eq:relationleftright}). $F_{{\rm L}}=-F_{{\rm R}}$, $ \phi^{{\rm L}}(x)=\phi^{{\rm R}}(-x)$.
The conjugated fields $\phi$, $\theta$ are customarily defined as
\begin{align}
\phi(x) &  \equiv  \frac{-\phi_{R}(x)+\phi_{L}(x)}{2},\nonumber \\
 & =-\frac{\pi x}{L}N_{R}-i\sum_{n>0}\frac{1}{\sqrt{n}}(a_{n}-a_{n}^{\dagger})\sin\left(\frac{n\pi x}{L}\right)e^{-\alpha\frac{\pi n}{2L}},\label{eq00005Cphi}\\
\theta(x) &  \equiv  \frac{\phi_{R}(x)+\phi_{L}(x)}{2},\nonumber \\
 & =\sum_{n>0}\frac{1}{\sqrt{n}}(a_{n}+a_{n}^{\dagger})\cos\left(\frac{n\pi x}{L}\right)e^{-\alpha\frac{\pi n}{2L}}.
\end{align}
The particle density operator $\rho$, which counts the number of
particle above the Fermi sea, is deduced from $\phi(x)$ through the
relation 
\begin{equation}
\rho(x)=-\frac{1}{\pi}\partial_{x}\phi(x).\label{eq:density}
\end{equation}
In the remaining, as we want to characterize the $0$ energy modes,
we will work at half-filling. It is important to notice that the definition
of the half-filling depends on the total number of sites. Let $N_F\in\mathbb{N}$ label the last occupied state of the Fermi sea. For $N$
even, we have $N_{F} \equiv  \frac{N}{2}$. The corresponding momentum is
$k_{F}=\frac{\pi N/2}{a(N+1)}\approx\frac{\pi}{2a}-\frac{\pi}{2aN}\approx\frac{\pi}{2a}-\frac{\pi}{2L}$
to first order in $1/L$. For $N$ odd, the two possible definitions
of the half-filled state are $N_{F}=\frac{N\pm1}{2}$ with corresponding
Fermi momenta $k_{F}=\frac{\pi}{2a}-\frac{\pi}{2L}(1\mp1)$. In the remaining of the paper, we will chose the convention $N_F=\frac{N-1}{2}$ for odd number of sites. The Fermi
sea state with all modes filled up to $N_{F}$ and empty above will
be referred to as the \emph{vacuum }state. 

We show in the SM \cite{SM} that $H$ can be expressed in terms of the
bosonic fields as 
\begin{align}
&H=  \int_{0}^{L}dx\bigg(\frac{v_{F}}{2\pi}\left(:(\partial_{x}\phi)^{2}:+:(\partial_{x}\theta)^{2}:\right)\label{eq:SSHbosonized}\\
 & +\frac{\delta}{L\sin\frac{\pi x}{L}}:\cos\left(2\phi(x)+\left(\frac{\pi}{a}-2k_{F}-\frac{\pi}{L}\right)x\right):\bigg)\nonumber 
\end{align}
where :: denotes normal-ordering of the fermionic modes with respect
to the vacuum and $v_{F} \equiv  2ta\sin(k_{F}a)$ the Fermi velocity. 
Formula (\ref{eq:SSHbosonized}) constitutes one of the main results of this
paper. We see that the SSH in the bosonized language is almost equivalent
to a sine-Gordon Hamiltonian except for the spatial dependence of the prefactor in front of the sinus. To derive (\ref{eq:SSHbosonized}),
we discarded constant terms and fast-varying modes $\propto e^{2ik_{F}x}$, so the bosonic
field describes modes with slow spatial variation with respect to the
lattice spacing. Since bosonization is a theory describing low energy
excitations, we also expect this expression to be valid for $\frac{\delta}{t}\ll1$. 

We now turn to the computation of $\rho$ using the classical Euler-Lagrange (EL) equations of motion. Let $Z \equiv  {\rm tr}(e^{-\beta H})$
and $\Pi$ the conjugated field to $\phi$, $\Pi \equiv  \frac{1}{\pi}\partial_{x}\theta.$
In the imaginary time formalism, we have 
\begin{align}
Z & =\int{\cal D}\phi{\cal D}\Pi e^{\int_{0}^{\beta}\int_{0}^{L}d\tau dx{\cal L}},
\end{align}
with ${\cal L}$ the Lagrangian density : $\int dx{\cal L}=\int dxi\Pi\partial_{\tau}\phi-H$. 

The EL equations $\frac{\partial{\cal L}}{\partial\phi}=\frac{\partial}{\partial\tau}\frac{\partial{\cal L}}{\partial\partial_{\tau}\phi}+\frac{\partial}{\partial x}\frac{\partial{\cal L}}{\partial\partial_{x}\phi}$
yields 
\begin{align}
 & \frac{1}{v_{F}}\partial_{\tau}^{2}\phi+v_{F}\partial_{x}^{2}\phi\nonumber \\
 & =-\frac{2\pi\delta}{L\sin\frac{\pi x}{L}}\sin\left(2\phi(x)+\left(\frac{\pi}{a}-2k_{F}-\frac{\pi}{L}\right)x\right).
\end{align}
In the $0$ temperature limit, all the weight of the probability measure
will be contained in the stationary solution $\partial_{\tau}\phi=0$. Introducing the natural rescaling
$y \equiv  x/L$, $\varphi(y) \equiv  \phi(yL)$, we get the $L$-independent equation 
\begin{align}
\partial_{y}^{2}\varphi(y) & =-\Delta\frac{\sin(2\varphi(y)+\epsilon\pi y)}{\sin(\pi y)}.\label{eq:ELeven}
\end{align}
where we introduced $\Delta=\frac{2\pi\delta L}{v_{F}}\approx\frac{\pi\delta(N+1)}{t}$
and $\epsilon=0$, for an \emph{even }number of sites where $k_{F}=\frac{\pi}{2a}-\frac{\pi}{2L}$
and $\epsilon=1$ for an \emph{odd }number of sites using the convention
$k_{F}=\frac{\pi}{2a}-\frac{\pi}{L}$. The boundary conditions for $\varphi$ are
read from Eq.(\ref{eq00005Cphi}): At $y=0$ we have $\varphi=0$
and at $y=1$ we have $\varphi=-\pi n_{{\rm R}}$ with $n_{{\rm R}}$
the number of particles created on top of the vacuum.

Let us make some remarks here. First, note that $\Delta$ is an adimensioned
parameter that fully characterizes the solution of the EL equations.
Note that $\Delta$ scales linearly in $N$ and $\frac{\delta}{t}$,
so increasing the system size has exactly the same effect as increasing
the ratio $\frac{\delta}{t}$. Interestingly, going from the odd to
the even case is equivalent to shift $\varphi$ by $\frac{\pi}{2}y$
which can be interpreted as substracting \emph{half} a particle to the system. 

\begin{figure}
\raggedright{}\includegraphics[width=0.95\columnwidth]{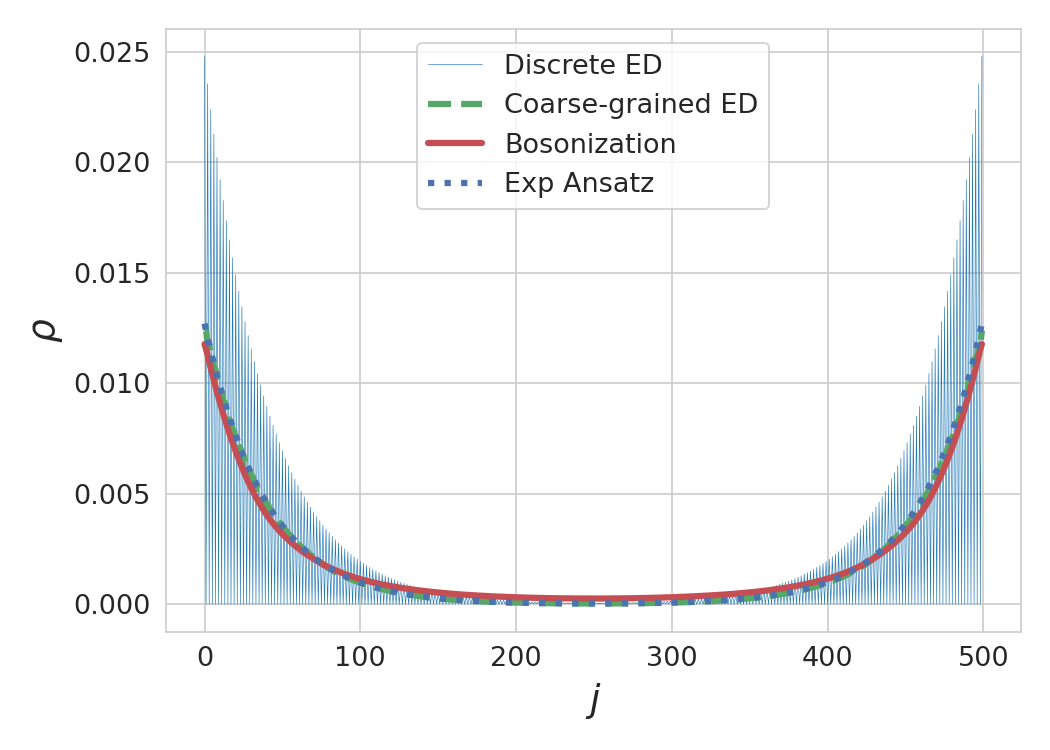}\caption{Comparison between the results of the discrete ED, the bosonization
result and the exponential fit for $\Delta=20$, $N=500$. The uniform vacuum density $\rho_j=1/2$ has been substracted. The light-blue
curve corresponds to exact discrete ED result which show fast oscillations
at the scale of the lattice spacing. The green dashed curve represents
the same data coarse-grained over $2$ sites. The red curve is
the density profile obtained by solving the EL equation (\ref{eq:ELeven})
from the bosonized theory and using $\rho_{j}=\frac{1}{2}-\frac{1}{\pi}\partial_{x}\phi(x=ja)$
and $\phi(x=ja)=\phi'(y)$. Lastly, the dashed blue line is obtained
from the exponential ansatz (\ref{eq:phiAnsatz}). Note that the latter two appear superposed on this plot.
\label{fig:edgemodes}}
\end{figure}

To the best of our knowledge, (\ref{eq:ELeven})
has no known analytical solution and we have to resort to numerics. To assert the validity of our approach, we compare numerical
solutions of (\ref{eq:ELeven}) with exact diagonalization
(ED) results performed on the discrete Hamiltonian (\ref{eq:discreteHamiltonian}) in the zero temperature ground state.
The ED results show fast oscillations on the scale of the lattice
spacing $\propto1/k_{F}$ that we do not see from the solutions of
the EL equations of motion since we precisely discarded these terms. Coarse-graining
over the fast oscillations gives a smoothly varying density profile
on the scale of the total system size. We observe excellent agreement
between the ED and the EL solutions - see Fig.\ref{fig:edgemodes}.
We also checked that the agreement holds both for $\delta$ positive
or negative, for an even or an odd number of sites and for different
values of $n_{{\rm R}}$ \cite{SM}. 

For the even case and $n_{R}=0$ the solution of the EL equations
is simply $\phi=0$ so that $\rho(x)=0$, which is consistent with the particle-hole
symmetry of the model. For $\delta>0$, fixing $n_{R}=1$ amounts to populate the
first mode above the vacuum state, i.e the Majorana edge mode. From
(\ref{eq:density},\ref{eq:ELeven}), we can thus deduce the density
profile of this edge mode. We plot on Fig.\ref{fig:edgemodes} the
numerical solution of (\ref{eq:ELeven}) and indeed see a concentration
of the density at the boundaries. There is a nice interpretation from
Eq.(\ref{eq:SSHbosonized}). The term proportional to $\delta$ wants
to lock the field in the minima of the $\cos$ term. For $N$ even
and $\delta>0$, this corresponds to $\phi=-\frac{\pi}{2}[\pi]$.
To match the boundary condition, the field $\phi$ must jump from
$0$ to $-\pi/2$ and then from $-\pi/2$ to $-\pi$. These jumps
translates in concentration on the edges for the density $\rho(x)=-\frac{1}{\pi}\partial_{x}\phi(x)$.
The stiffness of the jumps of $\phi$ is controlled by $\Delta$ and
determines how much the mode is concentrated at the edges. Eq.(\ref{eq:ELeven})
is consistent with the exponential localization of the edge mode near
the boundary. Indeed, for small $y$, one crude approximation of Eq.(\ref{eq:ELeven})
at first order in $y$ is given by $\partial_{y}(\partial_{y}\varphi)\approx-\frac{2\Delta}{\pi}\partial_{y}\varphi$.
Using additionally that the total particle number is 1, i.e $\frac{1}{\pi}\int_{0}^{1}dy\partial_{y}\varphi(y)=1$,
leads to the following ansatz $\varphi_{{\rm A}}$ when $\Delta\gg1$
: 

\begin{align}
\partial_{y}\varphi_{{\rm A}}(y) & =-\Delta\frac{\cosh(\frac{\Delta}{\pi}(2y-1))}{\sinh(\frac{\Delta}{\pi})}.\label{eq:phiAnsatz}
\end{align}
This exponential ansatz dictates the expression for the typical localization
length of the edge mode 
\begin{equation}
\ell \equiv  \frac{\pi L}{2\Delta}=\frac{v_{F}}{4\delta}.
\end{equation}
For the free SSH, it is known -see e.g \cite{AsbothBookintroductiontopologicalinsulator,CoursDalibardTopo}- that $\ell\propto\frac{a}{\ln\frac{t+\delta}{t-\delta}}\approx\frac{at}{2\delta}$
for $\frac{\delta}{t}\ll1$, which is consistent with our result since
$v_{F}\approx2ta$ at half-filling. 
Remark that being in the $\frac{\delta}{t}\ll1$ regime
automatically implies that $\ell\gg a$.
\begin{figure*}
\includegraphics[width=0.85\linewidth]{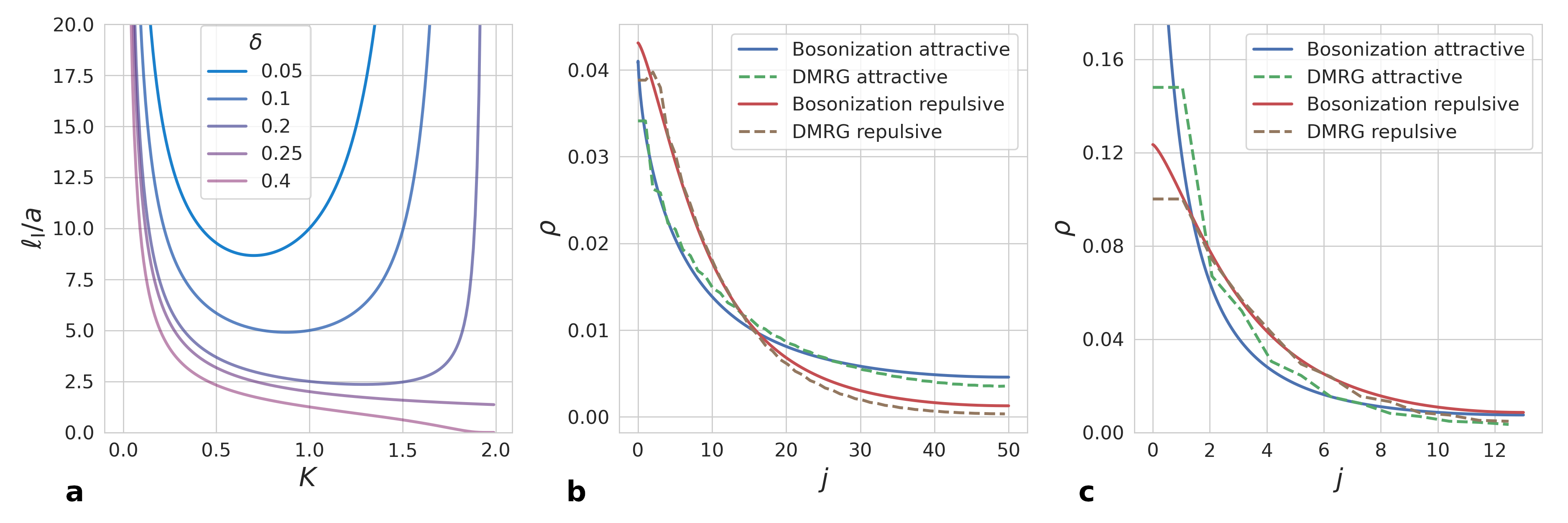}\caption{\textbf{a} Plots of relation \eqref{eq:localenghInteractions} for the localization length in the interacting case as a function of $K$ for different values of $\delta$. We see that the effect of interactions on the edge mode strongly depends on the value of $\delta$. \textbf{b} Comparison between the solution of the EL equations of motion (\ref{eq:ELEOM_INT})
in the interacting case with DMRG simulations. The DMRG results for attractive (repulsive) interactions have been coarse-grained once (twice) over two sites. Only half of the solution
is shown for better readibility. We took $N=100$, $\delta=0.05$,
$V=1.6$, $K=0.7$ for the repulsive case and $V=-0.77$, $K=1.4$
for the attractive case. In this case, attractive (repulsive) interactions delocalize (localize) the edge mode. \textbf{c} Same than b with parameters value $N=26$, $\delta=0.2$,
$V=1.6$, $K=0.7$ for the repulsive case and $V=-0.77$, $K=1.4$ for the attractive case. We see that, in comparison to case b, the qualitative effect of interactions are swaped.}
\label{fig:Tryptique}
\end{figure*}

We now turn to interactions. We consider a nearest neighbor interacting
term of the form 
\begin{equation}
H_{{\rm I}}:=V\sum_{j=1}^{N-1}\left(n_{j}-1/2\right)\left(n_{j+1}-1/2\right),\label{eq:interactingHamiltonian}
\end{equation}
with $n_{j}:=c_{j}^{\dagger}c_{j}$ the particle number operator.
From now on, we will work exclusively, in the topological phase, i.e
$\delta>0$, $N$ even and $N_{F}=\frac{N}{2}$. We show in the SM \cite{SM} that in the presence of this interacting term, the bosonization
procedure leads for the total Hamiltonian to 
\begin{align}
H= & \int_{0}^{L}dx\Big(\frac{1}{2\pi}\left(\frac{u}{K}:(\partial_{x}\phi)^{2}:_{{\rm I}}+uK:(\partial_{x}\theta)^{2}:_{{\rm I}}\right)\nonumber \\
 & +\delta\left(\frac{2}{a\pi}\right)^{1-K}\frac{1}{\left(L\sin\frac{\pi x}{L}\right)^{K}}:\cos(2\phi):_{{\rm I}}\Big).
\label{eq:Hamiltonianinteraction}
\end{align}
With $uK\equiv v_{F}$ and $\frac{u}{K}\equiv v_{F}+\frac{4Va}{\pi}$.
Since the free part of the Hamiltonian has been rescaled by interactions,
normal ordering needs to be done with respect to the new ``squeezed''
vacuum which we denote by $::_{{\rm I}}$. The $K$ dependence of
the prefactors of the $\cos$ terms is a direct consequence of that.
In principle, since we are at half-filling, there should also be a $\cos(4\phi)$ term \cite{SM}. For simplification, we will neglect this contribution in the present work as it is irrelevant in the RG sense if the interactions are not too repulsive, i.e if $K>1/2$. 
Eq. \eqref{eq:Hamiltonianinteraction} is the second crucial result of the paper.
The EL equations of motion in the presence of interactions become
\begin{equation}
\partial_{y}^{2}\varphi(y)=-2\left(\frac{L}{a}\right)^{2}\left(\frac{a\pi}{2L}\right)^{K}K^{2}\frac{\delta}{t}\frac{\sin(2\varphi)}{\left(\sin(\pi y)\right)^{K}}.\label{eq:ELEOM_INT}
\end{equation}
A comparison of numerical solutions of (\ref{eq:ELEOM_INT}) and density-matrix
renormalization group (DMRG) simulations is shown on Fig.\ref{fig:Tryptique}-\textbf{b} for $\delta=0.05$, $N=100$ and $V=1.6$, $K=1.4$
for the attractive case and $V=-0.77$, $K=1.4$ for the repulsive
case. We see that the EL equations of motion predicts the
correct density profile. For these parameters, we see that attractive
interactions delocalize the edge mode into the bulk while repulsive
interaction localize it further. 

 For $K<2$, we can give an estimation of the localization length by expanding (\ref{eq:ELEOM_INT}) for $y\ll1$. This gives 
\begin{equation}
\partial_{y}^{2}\varphi(y)=-\left(\frac{2L}{a}\right)^{2-K}K^{2}\frac{\delta}{t}\frac{\varphi(y)}{y^{K}}
\end{equation}
Imposing $\varphi(y)=0$, the solution to this equation are of the
form 
\begin{align}
\phi(x\ll L)=\sqrt{\frac{x}{L}}J_{\frac{1}{|2-K|}}\left(\left(\frac{2x}{a}\right)^{\frac{2-K}{2}}\left(\frac{\delta}{t}\right)^{\frac{1}{2}}\frac{2K}{|2-K|}\right)
\end{align}
where $J_{\alpha}$ is the Bessel function of the first kind - see \cite{SM} for the proof. The precise
shape of the Bessel function depends on $\alpha$ but, as one can easily
verify, the position of the first maximum of $J_\alpha$ scales linearly with
$\alpha$. Thus, we define the localization length to be the value
$\ell_{{\rm I}}$ such that $\varphi(\ell_{{\rm I}})=\frac{2}{|2-K|}$
where the factor $2$ has been put in order
to match with the localization length of the free case. Following this
definition, we obtain that 
\begin{equation}
\ell_{{\rm I}}=\frac{a}{2}\left(\frac{t}{\delta K^{2}}\right)^{\frac{1}{2-K}}. \label{eq:localenghInteractions}
\end{equation}
The localization length diverges at $K=2$ if $\frac{\delta}{t}<\frac{1}{4}$. This gives a rough criteria for having a localized mode in the attractive regime $K>1$.
Importantly, note that $\ell_{\rm I}$ is not, in general, a monotonic function of $K$ see Fig-\ref{fig:Tryptique}-\textbf{a}.
Interestingly, one consequence of this is that attractive
or repulsive interaction do not \emph{systematically }delocalize or
localize the edge mode, their effect can change depending on the value
of $\delta$. This is illustrated on Fig.\ref{fig:Tryptique}-\textbf{c} where we took $\delta=0.2$.
Contrary to the previous case shown on Fig.\ref{fig:Tryptique}-\textbf{b} , we see that the effects of attractive interactions is to \emph{localize} the edge mode further
to the boundary and the other way around for the repulsive ones.  

\textit{Conclusion} - In this paper, we derived the bosonized theory of the interacting SSH model with open boundaries. Importantly, our study offers quantitative arguments to determine the effects of interactions on the edge mode and pave the way to study other interacting topological models such as the spin 1 antiferromagnetic Heisenberg chain \cite{Haldanespin1chain1,Haldanespin1chain2}, the AKLT model \cite{AKLTmodel} or the Kitaev fermionic chain \cite{Kitaevchain}. One of our remarkable results for the SSH model is that attractive (repulsive) interactions do not systematically delocalize or localize the edge mode, but this behavior is strongly dependent on the value of the staggering parameter $\delta$.

We focused on the mean density profile of the edge mode but it would be interesting as a future direction to understand the interplay between interactions and quantum correlations of the edges. Another notable point is that we systematically discarded the Umklapp term in our study. Nevertheless, in the strongly repulsive regime, it is expected to play a role, leading to possibly interesting new phenomena.

\begin{acknowledgments}
\textbf{Acknowledgements} The DMRG simulations presented in this paper were done using the TeNPy package for tensor network calculations with python \cite{TeNPy}. T.J thanks Aashish Clerk for interesting discussions and comments on this work. The authors acknowledge support from the Swiss National Science Foundation under Division II.  \end{acknowledgments}

\bibliography{bibliobosonizationSSH}

\onecolumngrid
\appendix
\section*{Supplemental Material}
\section{Abacus}

For the reader's convenience, we first recall the expression of the
fermionic fields in terms of bosonic modes as well as various useful
relations they satisfy.

The continuous fermionic field $\Psi$ is split between a right-moving
field and a left-moving one : 

\begin{align}
\Psi(x) & =\Psi_{{\rm R}}(x)+\Psi_{{\rm L}}(x),\\
\Psi_{{\rm R}}(x) & =\sqrt{\frac{L}{2}}e^{ik_{F}x}\int_{-\infty}^{\infty}\frac{dk}{\pi}\tilde{c}_{k+k_{F}}e^{ikx},\\
\Psi_{{\rm L}}(x) & =-\Psi_{{\rm R}}(-x).\label{eq:relationleftright-1}
\end{align}
The main difference with bosonization on an infinite system size is
that the left mover is defined from the right mover for open boundaries.
The slow-mode is defined as
\begin{align}
{\rm R}(x) & \equiv F_{{\rm R}}\frac{1}{\sqrt{2\pi\alpha}}e^{i\phi^{{\rm R}}(x)},\\
\phi^{{\rm R}}(x) & \equiv \frac{\pi x}{L}N_{{\rm R}}+\sum_{n>0}\frac{1}{\sqrt{n}}(a_{n}e^{i\frac{n\pi x}{L}}+a_{n}^{\dagger}e^{-i\frac{n\pi x}{L}})e^{-\alpha\frac{\pi n}{2L}}.
\end{align}
One key observation is that ${\rm R}(x)$ can be interpreted as a
field with periodic boundary conditions on a system with size $2L$. 

The corresponding left mode is
\begin{align}
{\rm L}(x) & =F_{{\rm L}}\frac{1}{\sqrt{2\pi\alpha}}e^{i\phi^{{\rm L}}(x)},\\
F_{{\rm L}} & =-F_{{\rm R}},\\
\phi^{{\rm L}}(x) & =-\frac{\pi x}{L}N_{{\rm R}}+\sum_{n>0}\frac{1}{\sqrt{n}}(a_{n}e^{-i\frac{n\pi x}{L}}+a_{n}^{\dagger}e^{i\frac{n\pi x}{L}})e^{-\alpha\frac{\pi n}{2L}}.
\end{align}
and the canonical fields $\phi$ and $\theta$ are defined as 
\begin{align}
\phi(x) & :=\frac{-\phi_{\rm R}(x)+\phi_{\rm L}(x)}{2},\nonumber \\
 & =-\frac{\pi x}{L}N_{{\rm R}}-i\sum_{n>0}\frac{1}{\sqrt{n}}(a_{n}-a_{n}^{\dagger})\sin\left(\frac{n\pi x}{L}\right)e^{-\alpha\frac{\pi n}{2L}},\label{eq:=00005Cphi-1}\\
\theta(x) & :=\frac{\phi_{\rm R}(x)+\phi_{\rm L}(x)}{2},\nonumber \\
 & =\sum_{n>0}\frac{1}{\sqrt{n}}(a_{n}+a_{n}^{\dagger})\cos\left(\frac{n\pi x}{L}\right)e^{-\alpha\frac{\pi n}{2L}}.
\end{align}
It will turn out to be useful to divide the $\phi^{{\rm R}}$ operator
into an annihilation part $\varphi^{{\rm R}}$ and a creation part
with respect to the vacuum state : 
\begin{align}
\phi^{{\rm R}} & =\varphi^{{\rm R}}+\varphi^{{\rm R}\dagger},\\
\varphi^{{\rm R}} & =\frac{\pi x}{L}N_{R}+\sum_{n>0}\frac{1}{\sqrt{n}}a_{n}e^{i\frac{n\pi x}{L}}e^{-\alpha\frac{\pi n}{2L}},\\
\varphi^{{\rm R}\dagger} & =\sum_{n>0}\frac{1}{\sqrt{n}}a_{n}^{\dagger}e^{-i\frac{2n\pi x}{2L}}e^{-\alpha\frac{\pi n}{2L}}.
\end{align}
These operators have commutation relation 
\begin{align}
[\varphi^{R},\varphi^{R\dagger}] & =-\log(1-e^{-\alpha\frac{\pi}{L}}).
\end{align}
Using Baker-Campbell-Hausdorff relation $e^{A+B}=e^{A}e^{B}e^{-\frac{1}{2}[A,B]}$ when $[[A,B],B]=[[A,B],A]=0$
and $e^{-\frac{1}{2}[\varphi^{R},\varphi^{R\dagger}]}=\sqrt{1-e^{-\alpha\frac{\pi}{L}}}\approx\sqrt{\frac{\alpha\pi}{L}}$,
we can express $\mathrm{R}(x)$ in terms of normal-ordered operators
(Normal-ordering denoted by $::$ here refers to the normal order
of bosons with respect to the Fermi sea vacuum state denoted $\left|\Omega\right\rangle $)
: 
\begin{equation}
{\rm R}(x)=F_{\mathrm{R}}\frac{1}{\sqrt{2L}}:e^{i\phi^{\mathrm{R}}(x)}:.
\end{equation}
The normal ordered version of the left field is 
\begin{equation}
{\rm L}(x)=-F_{\mathrm{R}}\frac{1}{\sqrt{2L}}:e^{i\phi^{\mathrm{R}}(-x)}:.
\end{equation}
A useful relation allowing us to deal with product of normal ordered
field is 

\begin{equation}
:e^{ia\phi^{\mathrm{R}}(x)}::e^{ib\phi^{\mathrm{R}}(y)}:=e^{-ab\left\langle \Omega\right|\phi^{\mathrm{R}}(x)\phi^{\mathrm{R}}(y)\left|\Omega\right\rangle }:e^{ia\phi^{\mathrm{R}}(x)+ib\phi^{\mathrm{R}}(y)}:\label{eq:Normalorderingproduct}
\end{equation}
and an explicit calculation gets us 
\begin{equation}
e^{-ab\left\langle \Omega\right|\phi^{\mathrm{R}}(x)\phi^{\mathrm{R}}(y)\left|\Omega\right\rangle }=(1-e^{-\frac{\pi}{L}(\alpha+i(y-x))})^{ab}.
\end{equation}
Similarly, for the field $\phi$ :
\begin{equation}
:e^{ia\phi(x)}::e^{ib\phi(y)}:=:e^{ia\phi(x)+ib\phi(y)}:\left(\frac{\sin^{2}\left(\frac{\pi(x+y)}{2L}\right)}{\sin^{2}\left(\frac{\pi(x-y)}{2L}\right)}\right)^{-ab/4}. \label{eq:normalorderingphi}    
\end{equation}
Finally, another useful relation for normal-ordering is :
\begin{align}
e^{ib\phi(x)}&=:e^{ib\phi(x)}:e^{-\frac{b^2}{2}\left\langle \Omega\right|\phi^2(x)\left|\Omega\right\rangle},\\
&=:e^{ib\phi(x)}:\left(\frac{2L}{\pi\alpha}\sin\left(\frac{\pi x}{L}\right)\right)^{-\frac{b^{2}}{4}}.
\end{align}

\section{Derivation of the bosonized form of the SSH Hamiltonian}

In this part, we show how to derive the bosonized Hamiltonians (11)
and (18) of the main text.

The discrete Hamiltonian of the fermionic interacting SSH with open
boundaries was given by
\begin{equation}
H=\sum_{j=1}^{N-1}\left(-t-\delta(-1)^{j}\right)\left(c_{j}^{\dagger}c_{j+1}+{\rm h.c}\right)+V\left(n_{j}-\frac{1}{2}\right)\left(n_{j+1}-\frac{1}{2}\right).
\end{equation}
We split it in three parts that we will treat separately. $H=H_{0}+H_{1}+H_{{\rm I}}$
with $H_{0}$ the tight-binding term, $H_{1}$ the staggered part
and $H_{{\rm I}}$ the interacting part
\begin{align}
H_{0} & =-t\sum_{j=1}^{N-1}\left(c_{j}^{\dagger}c_{j+1}+{\rm h.c}.\right),\\
H_{1} & =-\delta(-1)^{j}\sum_{j=1}^{N-1}\left(c_{j}^{\dagger}c_{j+1}+{\rm h.c}.\right),\\
H_{{\rm I}} & =V\sum_{j=1}^{N-1}\left(n_{j}-\frac{1}{2}\right)\left(n_{j+1}-\frac{1}{2}\right).
\end{align}

\subsection{Bosonization of the tight-binding part $H_{0}$}

Although the bosonization of the tight-binding chain is a standard
textbook derivation, we redo it here because we are dealing with open
boundaries. Because of this, the left and right movers are not independent
fields and this may give rise to differences with the standard infinite
system case. 


Let us write $H_{0}$ in terms of the continuous fermionic field
:

\begin{align}
H_{0} & =-t\sum_{j=1}^{N-1}\left(c_{j}^{\dagger}c_{j+1}+{\rm h.c}.\right)\\
 & =-t\int_{0}^{L}dx\left(\Psi^{\dagger}(x)\Psi(x+a)+{\rm h.c}\right),\\
 & \approx-\frac{t}{2}\int_{-L}^{L}dx\left(e^{ik_{F}a}\mathrm{R}^{\dagger}(x)\mathrm{R}(x+a)+e^{-ik_{F}a}\mathrm{L}^{\dagger}(x)\mathrm{L}(x+a)+{\rm h.c}\right)
\end{align}
where going from the second line to the third we ``unfolded'' the
fields on $[-L,L]$ by making use of ${\rm R}(x)=-{\rm L}(-x)$ and
discarded fast oscillating terms proportional to $e^{\pm i2k_{F}x}$,
$\Psi_{\mathrm{R}}^{\dagger}\Psi_{\mathrm{L}}$ and $\Psi_{\mathrm{L}}^{\dagger}\Psi_{\mathrm{R}}$.
As usual in continuous field theory, the $a\to0$ limit must be regularized.
To this end, we denote by $::\mathrm{R}^{\dagger}(x)\mathrm{R}(x+a)::$
the \emph{point-splitting }procedure which consists in substracting
the infinite vacuum contribution to our operator

\begin{align}
::\mathrm{R}^{\dagger}(x)\mathrm{R}(x+a):: & =\lim_{a\to0}[\mathrm{R}^{\dagger}(x)\mathrm{R}(x+a)-\left\langle \Omega\right|\mathrm{R}^{\dagger}(x)\mathrm{R}(x+a)\left|\Omega\right\rangle ].
\end{align}

Using Eq.(\ref{eq:Normalorderingproduct}) we get 
\begin{align}
\mathrm{R}^{\dagger}(x)\mathrm{R}(x+a) & =\frac{1}{2L}:e^{-i\phi^{\mathrm{R}}(x)+i\phi^{\mathrm{R}}(x+a)}:(1-e^{-\frac{\pi}{L}(\alpha+ia)})^{-1},\\
\left\langle \Omega\right|\mathrm{R}^{\dagger}(x)\mathrm{R}(x+a)\left|\Omega\right\rangle  & =\frac{1}{2L}(1-e^{-\frac{\pi}{L}(\alpha+ia)})^{-1}.\\
::\mathrm{R}^{\dagger}(x)\mathrm{R}(x+a):: & =\frac{1}{2i\pi a}:\left(ia\partial_{x}\phi^{\mathrm{R}}(x)-\frac{1}{2}\left(a\partial_{x}\phi^{\mathrm{R}}(x)\right)^{2}+i\frac{a^{2}}{2}\partial_{x}^{2}\phi^{\mathrm{R}}(x)\right):
\end{align}
Similarly, for the left moving field, we have 

\begin{align}
::\mathrm{L}^{\dagger}(x)\mathrm{L}(x+a):: & =\frac{1}{2i\pi a}:\left(ia\partial_{x}\phi^{\mathrm{R}}(-x)+\frac{1}{2}\left(a\partial_{x}\phi^{\mathrm{R}}(-x)\right)^{2}-i\frac{a^{2}}{2}\partial_{x}^{2}\phi^{\mathrm{R}}(-x)\right):
\end{align}
When integrating these fields over $[-L,L]$, using the periodicity
of $\phi^{\mathrm{R}}$ on that interval, the terms $\partial_{x}\phi^{\mathrm{R}}$
and $\partial_{x}^{2}\phi^{\mathrm{R}}$ vanish. Now combining with
the h.c. we get
\begin{align}
-\frac{t}{2}\int_{-L}^{L}dx\left(e^{ik_{F}a}::\mathrm{R}^{\dagger}(x)\mathrm{R}(x+a)::+{\rm h.c.}\right) & =-i\frac{ta}{8\pi}\int_{-L}^{L}dx\left(e^{ik_{F}a}:\left(\partial_{x}\phi^{\mathrm{R}}(x)\right)^{2}:-e^{-ik_{F}a}:\left(\partial_{x}\phi^{\mathrm{R}}(x)\right)^{2}:\right),\\
 & =\underbrace{2ta\sin(k_{F}a)}_{=v_{F}}\frac{1}{8\pi}\int_{-L}^{L}dx:\left(\partial_{x}\phi^{\mathrm{R}}(x)\right)^{2}:.
\end{align}
And similarly for the left field 
\begin{align*}
-\frac{t}{2}\int_{-L}^{L}dx\left(e^{-ik_{F}a}::\mathrm{L}^{\dagger}(x)\mathrm{L}(x+a)::+{\rm h.c.}\right) & =\frac{v_{F}}{8\pi}\int_{-L}^{L}dx:\left(\partial_{x}\phi^{{\rm L}}(x)\right)^{2}:.
\end{align*}
Using $\phi^{{\rm R}}(x)=\phi^{{\rm L}}(-x)$ we can fold back everything
on the interval $[0,L]$. Using additionally that $(\partial_{x}\phi^{{\rm R}})^{2}+(\partial_{x}\phi^{{\rm L}})^{2}=2((\partial_{x}\phi)^{2}+(\partial_{x}\theta)^{2})$,
we end up with
\begin{equation}
H_{0}=\int_{0}^{L}dx\frac{v_{F}}{2\pi}\left(:(\partial_{x}\phi)^{2}:+:(\partial_{x}\theta)^{2}:\right).
\end{equation}

\subsection{Bosonization of the staggered part $H_{1}$. }

We now turn to the bosonization of the staggered part $H_{1}$. 
The philosophy is essentially the same as before except that, because
of the $(-1)^{j}$ prefactor, the slow-varying terms are going to
be the cross terms $\Psi_{\mathrm{R}}^{\dagger}\Psi_{\mathrm{L}}$
and $\Psi_{\mathrm{L}}^{\dagger}\Psi_{\mathrm{R}}$ and the fast varying
ones, that we will discard, $\Psi_{\mathrm{R}}^{\dagger}\Psi_{\mathrm{R}}$
and $\Psi_{\mathrm{L}}^{\dagger}\Psi_{\mathrm{L}}$. 
\begin{align}
H_{1} & =-\delta(-1)^{j}\sum_{j=1}^{N-1}\left(c_{j}^{\dagger}c_{j+1}+{\rm h.c}.\right),\\
 & =-\delta\int_{0}^{L}dxe^{i\frac{\pi x}{a}}\left(\Psi^{\dagger}(x)\Psi(x+a)+{\rm h.c}\right),\\
 & \approx-\delta\int_{0}^{L}dx\left(e^{i\frac{\pi x}{a}}e^{-i2k_{F}x}e^{-ik_{F}a}{\rm R}^{\dagger}(x){\rm L}(x+a)+e^{-i\frac{\pi x}{a}}e^{2ik_{F}x}e^{ik_{F}a}{\rm L}^{\dagger}(x){\rm R}(x+a)+{\rm h.c}\right)
\end{align}
where going from the second to the third line, we ignored the fast
oscillating terms and made use of the fact that $(-1)^{j}=e^{i\frac{\pi x}{a}}=e^{-i\frac{\pi x}{a}}$
for $x=ja$. Since we are half-filling $2k_{F}x\approx\frac{\pi x}{a}$
which will cancel the $(-1)^{j}$ contribution. However, as is discussed
in the main text, the expression of $k_{F}$ has subleading correction
in $1/L$ depending on the parity of the number of sites. For $N$
even, $k_{F}\approx\frac{\pi}{2a}-\frac{\pi}{2L}$ to first order
in $1/L$. For $N$ odd, the two possible Fermi momenta are $k_{F}\approx\frac{\pi}{2a}-\frac{\pi}{2L}(1\mp1)$
depending on the convention. We will use the convention $k_{F}=\frac{\pi}{2a}$
for odd sites in the following. It is important to keep track of $1/L$
term since $x$ itself ranges from $0$ to $L$. Thus, terms proportional to $x/L$
are in general not negligible in the thermodynamic limit. 

To regularize this expression, we again need to consider the point-splitting
of $H_{1}$. We focus on the first term : 

\begin{align}
::{\rm R}^{\dagger}(x){\rm L}(x+a):: & =\lim_{a\to0}[{\rm R}^{\dagger}(x){\rm L}(x+a)-\langle{\rm R}^{\dagger}(x){\rm L}(x+a)\rangle],\\
{\rm R}^{\dagger}(x){\rm L}(x+a) & =-\frac{1}{2L}:e^{-i\phi^{\mathrm{R}}(x)}::e^{i\phi^{\mathrm{R}}(-x-a)}:.
\end{align}
Using Eq.(\ref{eq:Normalorderingproduct}) we get
\begin{align}
{\rm R}^{\dagger}(x){\rm L}(x+a) & =-\frac{1}{2L}:e^{-i\phi^{\mathrm{R}}(x)+i\phi^{\mathrm{R}}(-x-a)}:(1-e^{-\frac{i\pi(-2x-a)}{L}})^{-1},\\
 & =-\frac{i}{4L\sin\frac{\pi(x+a/2)}{L}}e^{-\frac{i\pi(x+a/2)}{L}}:e^{2i\phi(x)+ia\partial_{x}\phi^{\mathrm{L}}(x)}:
\end{align}
which in the limit $a\to0$ becomes 
\begin{align}
{\rm R}^{\dagger}(x){\rm L}(x+a) & =-\frac{ie^{-\frac{i\pi x}{L}}}{4L\sin\frac{\pi x}{L}}:e^{2i\phi(x)}:
\end{align}
and
\begin{equation}
::{\rm R}^{\dagger}(x){\rm L}(x+a)::=-\frac{ie^{-\frac{i\pi x}{L}}}{4L\sin\frac{\pi x}{L}}(:e^{2i\phi(x)}:-1).
\end{equation}
Similarly 

\begin{align}
::{\rm L}^{\dagger}(x){\rm R}(x+a):: & =\frac{ie^{\frac{i\pi x}{L}}}{4L\sin\frac{\pi x}{L}}(:e^{-2i\phi(x)}:-1)\label{eq:pointsplittingSSH}
\end{align}
which leads to 
\begin{align}
H_{1} & =-\delta\int_{0}^{L}dx\frac{1}{4iL\sin\frac{\pi x}{L}}\left(e^{i\left(-k_{F}a+x(\frac{\pi}{a}-2k_{F}-\frac{\pi}{L})\right)}(:e^{2i\phi(x)}:-1)-e^{-i\left(-k_{F}a+x(\frac{\pi}{a}-2k_{F}-\frac{\pi}{L})\right)}(:e^{-2i\phi(x)}:-1)+{\rm h.c}\right),\\
 & =-\delta\int_{0}^{L}dx\frac{1}{L\sin\frac{\pi x}{L}}\left(:\sin\left(2\phi(x)+(\frac{\pi}{a}-2k_{F}-\frac{\pi}{L})x-k_{F}a\right):-\sin\left((\frac{\pi}{a}-2k_{F}-\frac{\pi}{L})x-k_{F}a\right)\right),\\
 & =\delta\int_{0}^{L}dx\frac{1}{L\sin\frac{\pi x}{L}}\left(:\cos\left(2\phi(x)+(\frac{\pi}{a}-2k_{F}-\frac{\pi}{L})x\right):-\cos\left((\frac{\pi}{a}-2k_{F}-\frac{\pi}{L})x\right)\right).
\end{align}
In the last line, we simplified the expression by making the approximation
$k_{F}a\approx\frac{\pi}{2}$. Note that such approximation cannot
be done for the $k_{F}x$ term. The part independent from $\phi(x)$
is precisely the contribution of the vacuum energy.
For the rest of the derivation, we will focus on the case of an even
number of sites for which $k_{F}=\frac{\pi}{2a}-\frac{\pi}{2L}$ so
that
\begin{align}
H_{1} & =\delta\int_{0}^{L}dx\frac{1}{L\sin\frac{\pi x}{L}}\left(:\cos\left(2\phi(x)\right):-1\right).
\end{align}

\subsection{Bosonization of the interacting part $H_{{\rm I}}$.}
We now derive the bosonization of the interacting term 
\begin{equation}
H_{{\rm I}}\equiv V\sum_{j=1}^{N-1}\left(n_{j}-\frac{1}{2}\right)\left(n_{j+1}-\frac{1}{2}\right)\label{eq:HinteracAPP-1}
\end{equation}
with $n_{j}$ the particle number operator $c_{j}^{\dagger}c_{j}$.
Recall
\begin{align*}
c_{j}^{\dagger}c_{j} & =a({\rm R}^{\dagger}(x){\rm R}(x)+{\rm L}^{\dagger}(x){\rm L}(x)+{\rm R}^{\dagger}(x){\rm L}(x)e^{-2ik_{F}x}+{\rm L}^{\dagger}(x){\rm R}(x)e^{2ik_{F}x}).
\end{align*}
The point-split expressions we previously found were 
\begin{align}
 & ::{\rm R}^{\dagger}(x){\rm R}(x)::=\frac{1}{2\pi}\left(:\partial_{x}\phi^{{\rm R}}:\right),\\
 & ::{\rm L}^{\dagger}(x){\rm L}(x)::=-\frac{1}{2\pi}\left(:\partial_{x}\phi^{{\rm L}}:\right),\\
 & ::{\rm R}^{\dagger}(x){\rm L}(x)::=-\frac{i}{4L}\frac{e^{-i\frac{\pi x}{L}}}{\sin\frac{\pi x}{L}}\left(:e^{2i\phi}:-1\right),\\
 & ::{\rm L}^{\dagger}(x){\rm R}(x)::=\frac{i}{4L}\frac{e^{i\frac{\pi x}{L}}}{\sin\frac{\pi x}{L}}\left(:e^{-2i\phi}:-1\right).
\end{align}
So that
\begin{align}
::{\rm R}^{\dagger}(x){\rm R}(x)::+::{\rm L}^{\dagger}(x){\rm L}(x):: & =-\frac{1}{\pi}\partial_{x}\phi,\\
::{\rm R}^{\dagger}(x){\rm L}(x)e^{-2ik_{F}x}::+::{\rm L}^{\dagger}(x){\rm R}(x)e^{2ik_{F}x}:: & =\frac{i}{4L}\frac{e^{-i\frac{\pi x}{a}}}{\sin\frac{\pi x}{L}}\left(:e^{2i\phi}:-:e^{-2i\phi}:\right),\\
 & =-\frac{e^{-i\frac{\pi x}{a}}}{\pi\alpha}\left(\sin(2\phi)\right).
\end{align}
Where we used the formula $e^{ib\phi(x)}=:e^{ib\phi(x)}:\left(\frac{2L}{\pi\alpha}\sin\left(\frac{\pi x}{L}\right)\right)^{-\frac{b^{2}}{4}}$. Thus,
\begin{align}
H_{{\rm I}}= & aV\int_{0}^{L}dx\left(-\frac{1}{\pi}\partial_{x}\phi(x)-\frac{e^{-i\frac{\pi x}{a}}}{\pi\alpha}\left(\sin(2\phi(x))\right)\right)\left(-\frac{1}{\pi}\partial_{x}\phi(x+a)-\frac{e^{-i\frac{\pi(x+a)}{a}}}{\pi\alpha}\left(\sin(2\phi(x+a))\right)\right).
\end{align}
Terms proportional to $e^{\pm ik_{F}x}$ in this product will be discarded
as they are oscillating rapidly. We will also discard any constant term
whenever they arise and systematically take the limit $a\to0$
when it is unambiguous. Note that, because we are at half-filling,
we have to keep the Umklapp terms proportional to $e^{\pm i4k_{F}x}$.
\begin{align}
H_{{\rm I}}= & aV\int_{0}^{L}dx\left(\frac{1}{\pi^{2}}\left(\partial_{x}\phi\right)^{2}-\frac{1}{(\pi\alpha)^{2}}\left(\sin(2\phi(x))\sin(2\phi(x+a))\right)\right),\\
= & aV\int_{0}^{L}dx\left(\frac{1}{\pi^{2}}\left(\partial_{x}\phi\right)^{2}+\frac{1}{2(\pi\alpha)^{2}}\left(\cos(4\phi(x))-\cos(2a\partial_{x}\phi)\right)\right),\\
\approx & aV\int_{0}^{L}dx\left(\frac{2}{\pi^{2}}\left(\partial_{x}\phi\right)^{2}+\frac{\cos(4\phi(x))}{2(\pi a)^{2}}\right)
\end{align}
where in the last line we identified the cut-off $\alpha$ and the
lattice spacing $a$. The normal-ordered version is 
\begin{equation}
H_{{\rm I}}=\int_{0}^{x}dx\left(\frac{2aV}{\pi^{2}}:(\partial_{x}\phi)^{2}:+\frac{V\pi^{2}a^{3}}{2^{5}\left(L\sin\frac{\pi x}{L}\right)^{4}}:\cos(4\phi):\right).
\end{equation}

\subsection{Putting everything together and normal-ordering with the new vacuum}

Putting everything together, we finally arrive for the total Hamiltonian
to 
\begin{equation}
H=\int_{0}^{L}dx\left(\frac{1}{2\pi}\left(\frac{u}{K}:(\partial_{x}\phi)^{2}:+uK:\left(\partial_{x}\theta\right)^{2}:\right)+\frac{\delta}{L\sin\frac{\pi x}{L}}:\cos(2\phi):+\frac{V\pi^{2}a^{3}}{2^{5}\left(L\sin\frac{\pi x}{L}\right)^{4}}:\cos(4\phi):\right),
\end{equation}
with $uK\equiv v_{F}$ and $\frac{u}{K}=v_{F}+\frac{4Va}{\pi}$. The
non-normal ordered version is 
\begin{equation}
H=\int_{0}^{L}dx\left(\frac{1}{2\pi}\left(\frac{u}{K}(\partial_{x}\phi)^{2}+uK\left(\partial_{x}\theta\right)^{2}\right)+\frac{2\delta}{a\pi}\cos(2\phi)+\frac{V}{2a\pi^{2}}\cos(4\phi)\right).
\end{equation}
With the interactions, the free part of the Hamiltonian has been renormalized.
We have to redefine the normal-ordering with respect to this new free
Hamiltonian in order to get the correct path integral formulation.
Let's call the new normal order $::_{{\rm I}}$. We introduce
the rescaled fields $\tilde{\phi}=\sqrt{K}\phi$ and $\tilde{\theta}=\frac{1}{\sqrt{K}}\theta$.
This transformation preserves the commutation relation between the
field operators. The free part of the Hamiltonian in terms of these
fields is expressed as 
\begin{equation}
\int_{0}^{L}\frac{dx}{2\pi}u\left((\partial_{x}\tilde{\phi})^{2}+(\partial_{x}\tilde{\theta})^{2}\right).
\end{equation}
We see that it has the same form than a genuine free Hamiltonian whose
Fermi velocity would be given by $u$. In particular, let $\left|\Omega_{{\rm I}}\right\rangle $
be the new vacuum. We have :
\begin{equation}
\left\langle \Omega_{{\rm I}}\right|\tilde{\phi}^{2}\left|\Omega_{{\rm I}}\right\rangle =\left\langle \Omega\right|\phi^{2}\left|\Omega\right\rangle =\frac{1}{2}\log\left(\frac{2L}{\pi\alpha}\sin(\frac{\pi x}{L})\right),
\end{equation}
so that 
\begin{align}
e^{i\zeta\phi} & =:e^{i\zeta\phi}:_{{\rm I}}e^{-\frac{\zeta^{2}}{2}K\left\langle \Omega_{{\rm I}}\right|\tilde{\phi}^{2}\left|\Omega_{{\rm I}}\right\rangle },\\
 & =:e^{i\zeta\phi}:_{{\rm I}}\left(\frac{2L}{\pi\alpha}\sin(\frac{\pi x}{L})\right)^{-\frac{K\zeta^{2}}{4}}.
\end{align}
The normal-ordered Hamiltonian thus becomes :
\begin{align}
H=&\int_{0}^{L}dx\Bigg(\frac{1}{2\pi}\left(\frac{u}{K}:(\partial_{x}\phi)^{2}:_{{\rm I}}+uK:(\partial_{x}\theta)^{2}:_{{\rm I}}\right)+\frac{2\delta}{a\pi}\left(\frac{\pi a}{2}\right)^{K}\frac{1}{\left(L\sin\frac{\pi x}{L}\right)^{K}}:\cos(2\phi):_{{\rm I}}\\
&+\frac{V}{2a\pi^{2}}\frac{1}{\left(\frac{2L}{\pi\alpha}\sin(\frac{\pi x}{L})\right)^{4K}}:\cos(4\phi):_{{\rm I}}\Bigg).
\end{align}
Note that in the main text, we neglected the Umklapp contribution which is the $\cos(4\phi)$ term. 

\section{Bessel functions}

In this section, we show that the differential equation
\begin{equation}
\partial_{x}^{2}h=-a\frac{h}{x^{b}}\label{eq:partialbessel}
\end{equation}
with $a,b$ real and positive admits as solution functions of the
form
\begin{equation}
h(x)\equiv\sqrt{x}f_{(|2-b|)^{-1}}\left(\sqrt{a}\frac{2}{|2-b|}x^{\frac{2-b}{2}}\right)
\end{equation}
where $f_{\alpha}$ is a Bessel function of the first or second kind,
i.e a function satisfying the following relation :
\begin{equation}
y^{2}f''_{\alpha}(y)+yf'_{\alpha}(y)+(y^{2}-\alpha^{2})f_{\alpha}(y)=0.
\end{equation}
Let $y=\mu x^{\nu}$ and $h(x)=\sqrt{x}f\left(y=\mu x^{\nu}\right)$
with $\mu\geq0$ and $\nu$ free variables for now. The previous equation
translates for $h$ into the relation 
\begin{align}
h(x)x^{-1/2}\left(\frac{1}{4\nu^{2}}-\alpha^{2}\right)+\frac{1}{\nu^{2}}h''(x)x^{3/2}+\mu^{2}x^{2\nu-1/2}h(x) & =0.
\end{align}
Fixing $\alpha=\frac{1}{|2\nu|}$, we end up with 
\begin{align}
h''(x) & =-\nu^{2}\mu^{2}x^{2(\nu-1)}h(x).
\end{align}
For $h$ to be solution of (\ref{eq:partialbessel}), we must have
\begin{align}
\nu & =\frac{2-b}{2},\\
|\mu| & =\sqrt{a}\frac{2}{|2-b|}
\end{align}
which ends our proof.


\section{Additional plots for the free case}

For the reader's convenience, we provide in this section additional
plots of the numerical solution of Eq.(14) of the main text as well
as comparison with ED result. We vary the sign of $\delta$ to go
from the topological to the trivial phase, the parity of the number
of sites $N$ and the number of excitations $n_{{\rm R}}$. 
For each plot, we substracted the average particle number occupation of the vacuum mode for the ED plots.

\begin{figure}[ht]
\includegraphics[width=1\textwidth]{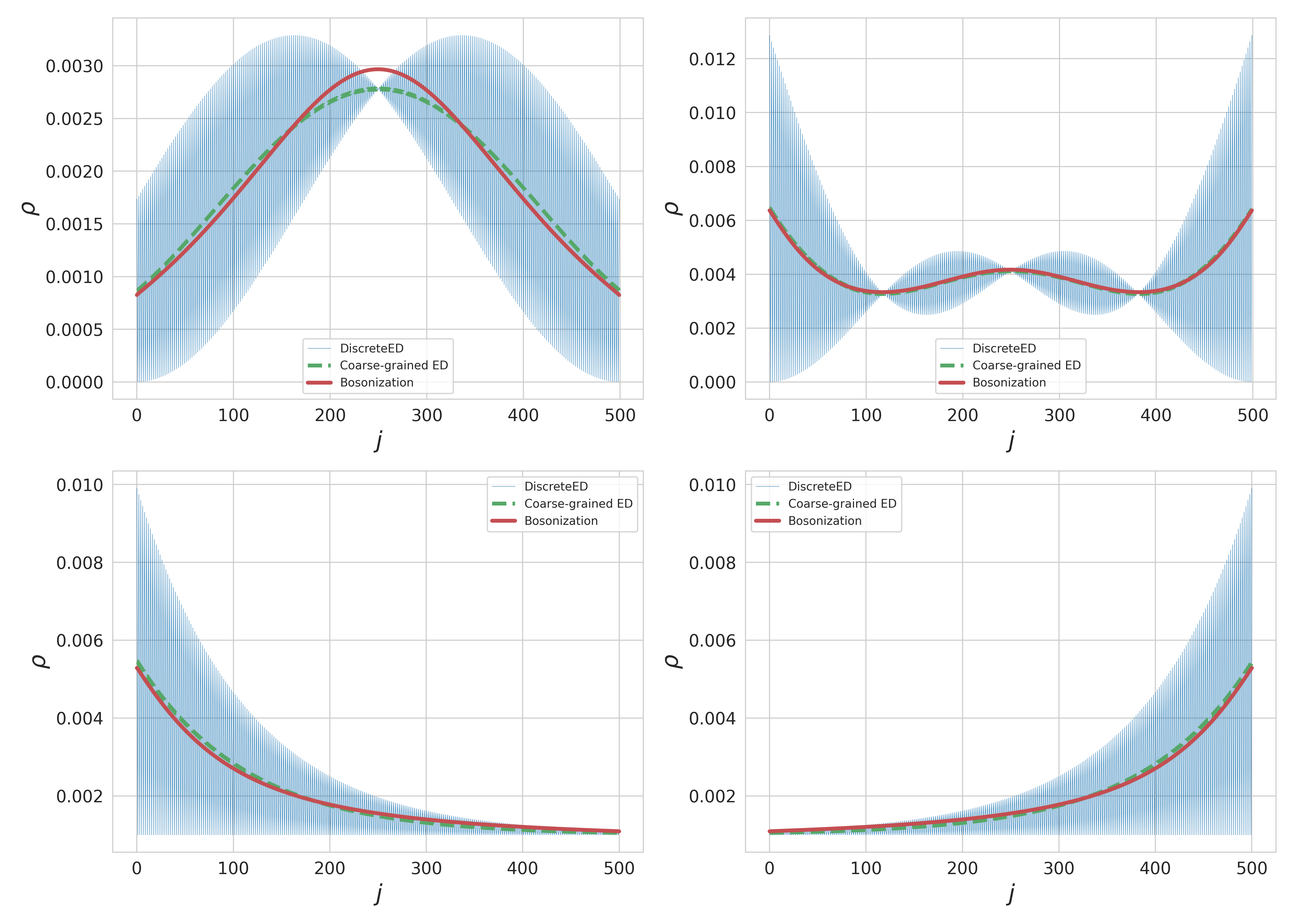}\caption{Additional plots comparing the solutions of the bosonization EL equations and the ED results. All the plots are done with $|\Delta|=7$ and $t=1$ \textbf{Top-Left :} First excited state of the trivial phase $N=500$, $\delta<0$, $n_{{\rm R}}=1$. \textbf{Top-Right :} Second excited state of the topological state $N=500$, $\delta>0$, $n_{{\rm R}}=2$. \textbf{Bottom-Left :} Odd number of sites $N=501$, $\delta>0$, $n_{{\rm R}}=1$. \textbf{Bottom-Right :} Odd number of sites $N=501$, $\delta<0$, $n_{{\rm R}}=1$. \label{fig:additionalplots}}
\end{figure}
\end{document}